\documentclass[aps,prb,amsmath,twocolumn]{revtex4}
\usepackage{bm}
\usepackage{graphicx}

\DeclareMathOperator{\Tr}{Tr}

\begin{document}

\title{Dephasing of Cooper pairs and subgap electron transport in superconducting hybrids}

\author{Andrew G. Semenov$^{1,2}$,
Andrei D. Zaikin$^{3,1}$
and Leonid S. Kuzmin$^{4,2}$}
\affiliation{$^1$I.E.Tamm Department of Theoretical Physics, P.N.Lebedev
Physics Institute, 119991 Moscow, Russia\\
$^2$Laboratory of Cryogenic Nanoelectronics, Nizhny Novgorod State Technical
University, 603950 Nizhny Novgorod, Russia\\
$^3$Institut f\"ur Nanotechnologie, Karlsruher Institut f\"ur Technologie
(KIT), 76021 Karlsruhe, Germany\\
$^4$Chalmers University of Technology, Goetenburg, Sweden
}

\begin{abstract}
We argue that electron-electron interactions fundamentally restrict 
the penetration length of Cooper pairs into a diffusive normal metal (N)
from a superconductor (S). At low temperatures this
Cooper pair dephasing length $L_\varphi$ remains finite and does not diverge at $T \to 0$. 
We evaluate the subgap conductance of NS hybrids in the presence of electron-electron 
interactions and demonstrate that this new length $L_\varphi$ can be directly extracted
from conductance measurements in such structures.

\end{abstract}

\maketitle

It is well known that a normal metal (N) attached to a superconductor (S)
also acquires superconducting properties \cite{dG,Tink}. This superconducting proximity
effect is directly related to the phenomenon of Andreev reflection \cite{And}: At the NS interface
Cooper pairs are converted into subgap quasiparticles (electrons) which can
diffuse deep into the normal metal keeping information about a macroscopic
phase of the superconducting condensate. Such macroscopic
quantum coherence of electrons in the normal metal gets destroyed by thermal
fluctuations only provided the corresponding inverse electron diffusion time
(Thouless energy) becomes smaller than temperature $T$. As a result, superconducting coherence
extends into a normal metal at a typical length $L_T \sim \sqrt{D/T}$ (where
$D$ is the electron diffusion coefficient) implying that the whole normal
metal can demonstrate superconducting properties at sufficiently low $T$.

This proximity induced superconductivity manifests itself in a number of
well known phenomena, such as Meissner and Josephson effects in
normal-superconducting hybrids \cite{bel,SaMiZhe} as well as
dissipative transport of subgap electrons across NS interfaces \cite{BTK}.
Provided the NS interface
transmission is low its corresponding subgap (Andreev) conductance
$G$ remains rather small being proportional to the second order in the
barrier transmission. On the other hand, $G$ can be strongly enhanced
at low energies due to non-trivial interplay between disorder and quantum
interference of electrons in the normal metal \cite{VZK,HN,HN1,Ben,Zai} which
leads to the so-called zero-bias anomaly (ZBA) $G \propto 1/\sqrt{V}$
and $G \propto 1/\sqrt{T}$ in the limit of low voltages and temperatures.

In this paper we will demonstrate that in the low temperature limit
both superconducting proximity effect and ZBA in Andreev conductance are
limited by {\it dephasing of Cooper pairs} due to electron-electron interactions
in the normal metal. Note that previously Coulomb effects in subgap electron
transport across NS interfaces were studied in a number of
works \cite{Zai,HHK,ZGalakt,ZGalakt2}, however decoherence effect of Coulomb interaction was
not yet addressed in a proper and complete manner. Below we will argue that
fluctuating electromagnetic field produced by fluctuating electrons in a disordered
normal metal destroys macroscopic coherence of electrons penetrating from a
superconductor at a typical length scale $L_{\varphi}$. The existence of
this length scale imposes fundamental limitations on the proximity effect
in NS hybrids at low temperatures $T \lesssim D/L_\varphi^2$. In
this temperature range the penetration depth
of superconducting correlations into the normal metal is not anymore given
by the thermal length $L_T$, but is limited by the dephasing length $L_\varphi$ which -- in
contrast to $L_T$ -- does not grow at $T \to 0$. We will evaluate Andreev
conductance $G$ for NS structures in the presence of electron-electron interactions
and demonstrate that in the low temperature limit $G$ essentially depends
on $L_\varphi$. This dependence allows to directly measure the dephasing
length $L_\varphi$ in transport experiments with NS hybrids.

It is also
interesting to point out that the dephasing length $L_\varphi$ derived here
for NS systems up to a numerical prefactor coincides with zero
temperature decoherence length obtained within totally different theoretical
framework \cite{GZ1,GZ2,GZ3,GZ4,GZ5} for a different physical quantity -- the so-called
weak localization (WL) correction to the normal metal conductance. This
agreement demonstrates fundamental
nature of low temperature dephasing by electron-electron interactions which
universally occurs in different types of disordered conductors, including
normal-superconducting hybrids.
On the other hand, as it will be explained further below, dephasing of Cooper pairs by
electron-electron interactions is in several important aspects different from
that for single electrons in a normal metal encountered, e.g., in the WL problem.

\begin{figure}[h]
\includegraphics[width=0.7\columnwidth]{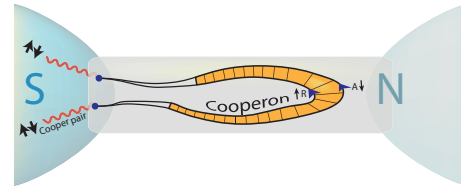}
\caption{(Color online) Hybrid SN structure under consideration and the
  diagram describing conversion of a Cooper pair into a pair of
electrons propagating inside the N-metal.}
\label{fig1}
\end{figure}

{\it The model and formalism.} Below we will analyze a hybrid SN structure
which consists of a normal metallic wire of cross-section
$a^2$ and length $L \gg a$  attached to bulk superconducting and normal
electrodes, as shown in Fig. 1.
The contact between the wire and the S-electrode is achieved via a small tunnel
barrier with cross-section $\Gamma$ and resistance
$R_I$ strongly exceeding the wire resistance $R_I\gg R=L/(\sigma a^2)$, where
$\sigma=2e^2\nu D$ is the wire Drude conductivity, $e$ is the electron charge
and $\nu$ is the density of states per spin direction.

In order to proceed we will employ the Keldysh
version of the nonlinear $\sigma$-model \cite{KA,KA1} adapted to SN structures
\cite{ZGalakt}. The effective action for our system
defined on the Keldysh contour with forward (F) and backward (B) parts consists of two terms
$S=S_w+S_{\Gamma}$ describing respectively diffusive motion of electrons
in the wire,
\begin{equation}
S_w[\check Q,{\bf A},\Phi]=\frac{i\pi\nu}{4}\Tr[D(\check\partial \check Q)^2-4\check\Xi\partial_t\check Q+4i\check\Phi\check Q],
\label{swire}
\end{equation}
and tunneling between the wire and the superconductor,
\begin{equation}
S_\Gamma[\check Q,{\bf A},\Phi]=-\frac{i\pi}{4e^2
  R_{I}\Gamma}\Tr_\Gamma[\check Q_{\rm sc},\check Q],
\label{sI}
\end{equation}
where $\check Q_{\rm sc}$ and $\check Q$ are taken at superconducting and normal
sides of the insulating barrier, $[x,y]$ denotes the commutator and "$\Tr$"\  implies the 
trace over the matrix indices as well as the integration over times
and coordinates. The covariant derivative is defined as
\begin{equation}
\check\partial \check Q=\partial_{{\bf r}}\check Q-i[\check\Xi\check {\bf
A},\check Q], \quad \check\Xi=\left(\begin{array}{cc} \hat \sigma_z &  0 \\
0 & \hat\sigma_z \end{array}\right).
\end{equation}
Here and below $ \hat\sigma_{x,y,z}$ denotes the set of Pauli matrices. 
Both parts of the action (\ref{swire}) and
(\ref{sI}) depend on the $4\times4$ dynamical matrix field $\check Q$ satisfying the
normalization condition $\check Q^2 =\check 1\delta(t-t')$
as well as on the fluctuating scalar and vector potentials
$\Phi({\bf r},t)$ and ${\bf A}({\bf
  r},t)$ which are defined on the Keldysh
contour and which account for the effect of electron-electron interactions.
We define
$\Phi^{\pm}=\frac{1}{\sqrt{2}}(\Phi^F\pm\Phi^B)$ and  ${\bf
  A}^{\pm}=\frac{1}{\sqrt{2}}({\bf A}^F\pm{\bf A}^B)$ and introduce the matrices
\begin{equation}
\check\Phi =\left(\begin{array}{cc} \Phi^+\hat1 & \Phi^-\hat1  \\
 \Phi^-\hat1 & \Phi^+ \hat1
 \end{array}\right), \quad\check{\bf A}=\left(\begin{array}{cc} {\bf A}^+\hat1 & {\bf A}^-\hat1  \\
 {\bf A}^-\hat1 & {\bf A}^+ \hat1
 \end{array}\right).
\end{equation}

{\it Perturbation theory and Gaussian integration.} In what follows we will restrict our consideration to
energies well below the superconducting gap and set
\begin{equation}
\check Q_{\rm sc}(t,t')=\left(\begin{array}{cc}\hat \sigma_y &
0\\0&\hat\sigma_y\end{array}\right)\delta(t-t').
\end{equation}
We will employ the so-called $\mathcal K$-gauge trick\cite{KA,KA1} which amounts to performing the gauge transformation
$\check Q({\bf r},t,t')\to e^{i\check\Xi \check{\mathcal K}({\bf r},t)}\check Q({\bf r},t,t') e^{-i\check\Xi \check{\mathcal K}({\bf r},t')}$
in order to eliminate linear terms in both electromagnetic potentials and deviations from the N-metal saddle point
\begin{gather}
\check Q_N=\check {\mathcal U}\circ \left(\begin{array}{cc}
\hat\sigma_z&0\\0&-\hat\sigma_z
\end{array}\right)\check {\mathcal U},\\
%\end{equation}
%\begin{equation}
\check{\mathcal U}(t-t')=\left(\begin{array}{cc}\delta(t-t'-0)\hat 1 &
-\frac{iT}{\sinh(\pi T(t-t'))}\hat 1 \\ 0 &
-\delta(t-t'+0)\hat1\end{array}\right).
\end{gather}
This goal is accomplished with the choice of the $\mathcal K$-field obeying the following equations
%\begin{multline}
\begin{eqnarray}
 \Phi_{\mathcal K}^+({\bf r},t)&=&D\partial_{\bf r}{\bf A}_{\mathcal K}^+({\bf r},t) \\
&&-2iDT\int dt'\coth(\pi T(t-t'))\partial_{\bf r}{\bf A}_{\mathcal K}^-({\bf r},t'),\nonumber\\
%\end{multline}
%\begin{equation}
 \Phi_{\mathcal K}^-({\bf r},t)&=&-D\partial_{\bf r}{\bf A}_{\mathcal K}^-({\bf r},t)
\end{eqnarray}
with $\Phi_{\mathcal K}({\bf r},t)=\Phi({\bf r},t)-\partial_t\mathcal K({\bf r},t)$ and ${\bf A}_{\mathcal K}({\bf r},t)={\bf A}({\bf
r},t)-\partial_{\bf r}\mathcal K({\bf r},t)$. After this transformation the action retains its initial form if one substitutes 
$  \check Q_{\rm sc}(t,t')\to e^{-i\check\Xi \check{\mathcal K}({\bf r},t)}\check Q_{\rm sc}(t,t')e^{i\check\Xi \check{\mathcal K}({\bf r},t')}$, $\Phi\to\Phi_{\mathcal K}$ and ${\bf A}\to{\bf A}_{\mathcal K}$.

Treating the tunneling term (\ref{sI}) perturbatively and performing the integration over the $\check Q$-field, similarly to \cite{ZGalakt} we 
arrive at the Andreev contribution to our action
\begin{equation}
S_A=-\frac{i}{32}\left(\frac{\pi}{e^2R_I\Gamma}\right)^2\langle \Tr_\Gamma[\check Q_{\rm sc},\check Q]\Tr_\Gamma[\check Q_{\rm sc},\check Q]\rangle_Q.
\label{sa}
\end{equation}
The dependence of this term on the electromagnetic potentials is encoded both in $\check Q_{\rm sc}$ and in the average of the $\check Q$-fields. 
Evaluating $S_A$ within the Gaussian approximation we will employ the parametrization \cite{KA,KA1}
%\begin{equation}
$
 \check Q\approx\check Q_0
 +i\check Q_0\circ {\mathcal U}\circ\check { W}\circ {\mathcal U}-\frac12\check Q_0\circ\check {\mathcal U}\circ\check { W}\circ\check { W}\circ\check {\mathcal U}
$
%\end{equation}
with
\begin{equation}
\check W =\left(\begin{array}{cccc} 0 &c_1 ({\bf r},t,t') & d_1 ({\bf r},t,t') & 0 \\
\bar c_1 ({\bf r},t',t)& 0& 0 & d_2 ({\bf r},t,t') \\
\bar d_1 ({\bf r},t',t)& 0& 0 & c_2 ({\bf r},t,t') \\
0&\bar d_2 ({\bf r},t',t)&  \bar c_2 ({\bf r},t',t) & 0
\end{array}\right).\nonumber
\end{equation}
Here $d_i$ and $c_i$ are respectively the diffuson and the Cooperon fields. Expanding the action $S_w$ up to the second order in these fields one recovers 
four different contributions $S_w=S_w^{(0,2)}+S_w^{(1,2)}+S_w^{(2,1)}+S_w^{(2,2)}$, where $S^{(i,j)}$ contains $i$-th power of the electromagnetic 
potentials and $j$-th power of $\check W$. By direct calculation one can verify that the term $S^{(2,1)}$ depends only on the diffuson fields 
which are irrelevant for the problem considered here. Hence, our action does not contain the first power of the Cooperon fields, and the corresponding 
propagator -- the Cooperon $\mathcal C$ -- can be obtained as a solution of a linear inhomogeneous equation containing the first and the second powers of 
electromagnetic potentials. 

At this stage we would like to remark that the spin structure of the Cooperon analyzed here differs from that of the Cooperon encountered, 
e.g., in the weak localization (WL) problem in disordered normal metals.  Indeed, representing the Cooperon as a sum of impurity ladder diagrams 
involving retarded ($G^R$) and advanced ($G^A$) Green functions one observes that the spin structure of the Cooperon responsible for the WL 
correction to the N-metal conductance is either $(\uparrow\uparrow)$ or $(\downarrow\downarrow)$ implying that both $G^R$ and $G^A$
correspond to either spin up or spin down states. 
In contrast, the spin structure of the Cooperon relevant for the proximity induced superconductivity is either $(\uparrow\downarrow)$ or $(\downarrow\uparrow)$ 
simply because Cooper pairs are spin-singlets. It is straightforward to verify that only the Cooperon formed by antisymmetric combination
$(\uparrow\downarrow-\downarrow\uparrow)/\sqrt{2}$ contributes to the subgap Andreev conductance of NS structures. In the presence 
of electron-electron interactions this Cooperon differs from that corresponding to other possible spin configurations
already on the level of the first order perturbation theory.

\begin{figure}[h]
\includegraphics[width=0.95\columnwidth]{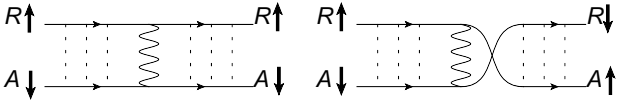}
\caption{First order interaction corrections for the Cooperon. The dashed lines represent scattering on impurities and the wavy line accounts for electron-electron interactions.}
\label{fig2}
\end{figure}

For illustration let us consider the first order diagrams depicted in Fig. 2. While in each of the cases 
$(\uparrow\uparrow)$, $(\downarrow\downarrow)$ and $(\uparrow\downarrow+\downarrow\uparrow)/\sqrt{2}$ these diagrams 
cancel each other exactly at $T=0$ and in the limit of zero frequencies and momentum, no such cancellation occurs
for the antisymmetric combination $(\uparrow\downarrow-\downarrow\uparrow)/\sqrt{2}$
because of extra minus sign encountered in this case.
Thus, for the latter spin combination (which is only relevant here) non-vanishing zero temperature dephasing 
is observed already within the first order perturbation theory in the interaction. 

{\it Andreev conductance.} Below we will proceed non-perturbatively and evaluate the subgap Andreev current $I$ to all orders in the interaction.
Defining this current as
\begin{equation}
 I= \frac{e}{2}\int_\Gamma d^{2}{\bf r}\langle\delta S_A/\delta {\mathcal K}^-({\bf r})\rangle_\Phi 
\label{Ac}
\end{equation}
and calculating $S_A$ along the lines with the analysis \cite{ZGalakt} which now includes $\Phi^+_{\mathcal K}$ and ${\bf A}^+_{\mathcal
K}$, from Eq. (\ref{Ac}) we obtain
\begin{equation}
I=\frac{\pi T}{2\nu e^3(R_I\Gamma)^2}\int_\Gamma d^{2}{\bf
r}d^{2}{\bf r}'\int d\tau {\rm Im}\frac{\langle\mathcal P({\bf r},{\bf r}',\tau;t) e^{ieV\tau}\rangle_\Phi}{\sinh(\pi T\tau)},
\label{Ap}
\end{equation}
where $\mathcal P=\mathcal C({\bf r},\tau;{\bf r}',0;t-\tau/2)e^{-2i\mathcal K^+({\bf r},t)+2i\mathcal K^+({\bf r}',t-\tau/2)}$ 
with the Cooperon $\mathcal C$  obeying the  equation
\begin{widetext}
\begin{multline}
\left( 2\partial_\tau-i\Phi_{\mathcal K}^+({\bf r},T-\tau/2)+i\Phi_{\mathcal K}^+({\bf r},T+\tau/2)-D(\partial_{\bf r}+i{\bf A}^+_{\mathcal K}({\bf r},T-\tau/2)+i{\bf A}^+_{\mathcal K}({\bf r},T+\tau/2))^2 \right)\mathcal C({\bf r},\tau;{\bf r}',\tau';T)\\=\delta({\bf r-r'})\delta(\tau-\tau')
\label{coop}
\end{multline}
\end{widetext}
and $V$ is an external voltage bias. Note that here and below we keep only the fields $\Phi^+$ and $\mathcal K^+$ neglecting  
$\Phi^-$ and $\mathcal K^-$ which are irrelevant for dephasing of Cooper pairs \cite{FN}. 

Resolving Eq. (\ref{coop}), we get
\begin{widetext}
\begin{multline}
\mathcal P({\bf r},{\bf r}',\tau;t)=
\frac{\theta(\tau) e^{i\mathcal K^+({\bf r},t-\tau)-i\mathcal K^+({\bf r},t)} }{2}\int\limits_{{\bf x}(0)={\bf r}'}^{{\bf x}(\tau)={\bf r}}\mathcal D{\bf x}e^{-\int\limits_{0}^\tau dt'
\left(\frac{(\dot{\bf x}(t'))^2}{2D}-\frac{i}{2}(\Phi^+({\bf x}(t'),t-(t'+\tau)/2)-\Phi^+({\bf x}(t'),t+(t'-\tau)/2))\right)}.
\end{multline}
\end{widetext}
What remains is to perform a straightforward Gaussian average over
$\Phi^+$-fields as well as an average over diffusive trajectories. The latter
average is handled approximately with the aid of the formula
$\langle e^{F}\rangle_{\rm diff}\simeq e^{\langle F\rangle_{\rm diff}}$. As a result we find
\begin{equation}
 \langle\mathcal P({\bf r},{\bf r}',\tau;t)\rangle_\Phi=\mathcal D({\bf r},{\bf r}';\tau)e^{-f({\bf r},{\bf r}',\tau)}
\label{pav}
\end{equation}
with $f({\bf r},{\bf r}',\tau)=f_t({\bf r},\tau)+f_b({\bf r},{\bf r}',\tau)+f_{tb}({\bf r},{\bf r}',\tau)$,
\begin{eqnarray}
\label{ft}   
f_t({\bf r},\tau)=\frac{i}2
\left(\mathcal V_{\mathcal K\mathcal K}^{++}({\bf r},{\bf r},0)-\mathcal V_{\mathcal K\mathcal K}^{++}({\bf r},{\bf r},\tau)\right),\\
%\begin{multline}
%    f_{tb}({\bf r},{\bf r}',\tau)=i\int\limits_0^\tau dt\int d^{d}{\bf x}\frac{\mathcal D({\bf r},{\bf x};\tau-t)\mathcal D({\bf x},{\bf r}';t)}{\mathcal D({\bf r},{\bf r}';\tau)}\\\times\left(\mathcal V_{\mathcal K\Phi}^{++}({\bf r},{\bf x},(\tau-t)/2)- \mathcal V_{\mathcal K\Phi}^{++}({\bf r},{\bf x},(\tau+t)/2)\right),
%\label{ftb}
%\end{multline}
%\begin{multline}
f_b({\bf r},{\bf r}',\tau)=i\int\limits_{0}^\tau dt\int\limits_{0}^tdt'\int d^d{\bf x}d^d{\bf x}'\nonumber \\
\times\left(\mathcal V_{\Phi\Phi}^{++}({\bf x},{\bf x}',(t-t')/2)- \mathcal V_{\Phi\Phi}^{++}({\bf x},{\bf x}',(t+t')/2)\right)\nonumber\\
\times
\frac{\mathcal D({\bf r},{\bf x};\tau-t)\mathcal D({\bf x},{\bf x}';t-t')\mathcal D({\bf x}',{\bf r}';t')}{\mathcal D({\bf r},{\bf r}';\tau)},
\label{fb}
\end{eqnarray}
where $\mathcal D({\bf x},{\bf x}',t)$ is the diffusive propagator,  
$\mathcal V^{++}_{\Phi\Phi}({\bf r},{\bf r}',t-t')=-2i\langle\Phi^+({\bf r},t)\Phi^+({\bf r}',t')\rangle_\Phi$, and $\mathcal V^{++}_{\mathcal K\Phi}$, $\mathcal V^{++}_{\mathcal K\mathcal K}$ is defined analogously.
The function $f_{tb}$ is expressed via the correlator $\langle{\mathcal K}^+\Phi^+\rangle$. We chose to omit it here since  
$f_{tb}$ remains much smaller than both $f_t$ and $f_b$. 

Eqs. (\ref{Ap}), (\ref{pav})-(\ref{fb}) define the central result of our work which describes the effect of 
electron-electron interactions on the subgap current in diffusive NS structures.

{\it Quasi-1d structures.} Below we will concentrate on quasi-1d $N$-metal wires (Fig. 1) and set $\Gamma =a^2$. In this case 
the differential Andreev conductance $G(V)=dI/dV$ takes the form 
\begin{eqnarray}
G=\frac{\pi T}{4\nu e^2 R_I^2}\int\limits_0^\infty d\tau^2\frac{\mathcal D(0,0;\tau)\cos(eV\tau)}{\sinh(\pi T\tau)}e^{-f(0,0,\tau)}
\label{finres}
\end{eqnarray}
with $\mathcal D(0,0;\tau)=\vartheta_2(0,e^{-\tau/\tau_D})/(2La^2)$, where $\vartheta_2$ is the second Jacobi theta-function and $\tau_D=2L^2/(\pi^2D)$ 
is the Thouless time. The function $f$ 
accounts for dephasing of Cooper pairs. For $\pi T\tau\ll 1$ Eqs. (\ref{ft}), (\ref{fb}) yield
\begin{multline}
f(0,0,\tau)\simeq\frac{8}{g}\ln\left(\frac{\tau}{\tau_{RC}}\right)+\frac{\tau}{\tau_{\varphi}}
+\sqrt{\frac{\pi\tau\tau_c}{4\tau_{\varphi}^2}}\ln\left(\frac{\tau_c}{\tau}\right).
\label{flim}
\end{multline}
%\begin{multline}
%f_t+f_b\approx\frac{8}{g}\ln\left(\frac{\sinh(\pi T\tau/2)}{2T\tau_c}\right)+\frac{\tau}{\tau_{\varphi}}
%\\+\sqrt{\frac{\pi\tau\tau_c}{4\tau_{\varphi}^2}}\left(\ln\left(\frac{\tau_c}{\tau}\right)-7.541\right)
%+10.604\frac{\tau_c}{\tau_{\varphi}}
%\end{multline}
In the first term in Eq. (\ref{flim}) we defined dimensionless conductance $g=4\pi\nu Da^2/L \gg 1$ and $\tau_{RC}=RC$,
where $C$ is an effective capacitance. This term is caused by spatially uniform fluctuations of the scalar potential
and matches with the results \cite{HHK,ZGalakt}.  
The remaining terms in Eq. (\ref{flim}) originate from
non-uniform in space fluctuations in the N-metal and
define the new scales in our problem -- Cooper pair decoherence time $\tau_\varphi=2\pi\nu a^2\sqrt{2D\tau_c}$
and decoherence length $L_\varphi =\sqrt{D\tau_\varphi}$, where $\tau_c \sim l/v_F$ sets a short time cutoff \cite{GZ1,GZ3,GZ4}
and also $\tau_\varphi \gg \tau_{RC}$.
Note that up to an unimportant prefactor of order one  $\tau_\varphi$ coincides with zero temperature
electron decoherence time evaluated, e.g., for the WL problem \cite{GZ1,GZ3,GZ4}.

At this point we would like to emphasize that the agreement between the low temperature dephasing length scales $L_\varphi$ found here
for Cooper pairs and previously \cite{GZ1,GZ2,GZ3,GZ4,GZ5} for single electrons is by no means a pure coincidence. 
Rather this agreement reflects fundamental and universal nature of low temperature quantum decoherence caused by electron-electron 
interactions in various types disordered conductors. At the same time, the Cooperon encountered in the WL problem is in 
many respects different -- both qualitatively and quantitatively -- from that studied here. As we already indicated above, 
the most important difference is that the spin structure of our Cooperon (antisymmetric combination
of spin-singlets) corresponds to that of a Cooper pair and is entirely different
from that for the Cooperon in the WL problem. In addition, the Cooperon describing 
propagating Cooper pairs in the normal metal is naturally bound to the NS interface,
which is obviously not the case in the WL problem. As a result these two Cooperons are defined by 
formally different diagrammatic series and, hence, no apriori conclusions could possibly be drawn 
for our present problem from the Cooperon analysis developed for single electrons in disordered metals.

These differences have several important implications.
For clarity, let us summarize the most important ones again:
($i$) Unlike single electrons in normal metals, Cooper pairs in NS structures 
get dephased already by {\it uniform} fluctuations of the scalar potential, as described by the first term in Eq. (\ref{flim}), 
($ii$) unlike in case of the WL problem, non-vanishing dephasing of Cooper pairs at
$T=0$ occurs already within the {\it first order} perturbation theory in the electron-electron interactions, see
Fig. 2 and the corresponding discussion above,
($iii$) already at $T=0$ the
Cooperon studied here decays differently as compared to the Cooperon in the WL
problem, cf., e.g.,
our Eq. (19) and Eq. (28) in \cite{GZ4}, and ($iv$) at not very low $T$ the temperature
dependent decay time for the Cooperon in NS systems is entirely different from that 
in the WL problem \cite{FN0}.

\begin{figure}[h]
\includegraphics[width=0.99\columnwidth]{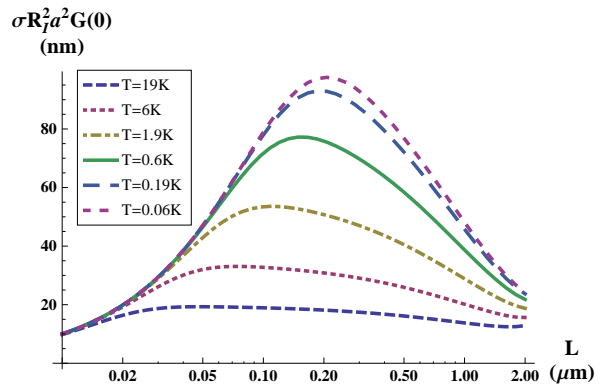}
\caption{(Color online) $G(0)$ as a function
of $L$ for $a=10$ nm, $D=21$ cm$^2$/s. For these parameter values one finds 
$1/\tau_\varphi \sim 0.6$ K and $L_\varphi \sim 0.2$ $\mu$m.} \label{fig3}
\end{figure}

Turning to concrete results we first consider the low voltage limit $eV \ll T$.
At high temperatures $T\gg 1/\tau_\varphi$ the penetration length of Cooper pairs into the N-metal is
defined by $L_T$, while $L_\varphi$ is irrelevant and dephasing is only due to spatially uniform fluctuations
described by the first term in Eq. (\ref{flim}). In this case the results \cite{ZGalakt} are reproduced and 
one finds $G(0) \propto T^{8/g-1/2}$.  At low temperatures $T\ll 1/\tau_\varphi$, on the contrary, $L_T$ becomes irrelevant and the penetration length
of superconducting correlations into the N-metal is set by $L_\varphi$. Then for the linear subgap conductance we obtain
\begin{equation}
G(0)\simeq\begin{cases} \frac{1}{\sigma
R_I^2a^2}\left(\frac{4\tau_{RC}}{\tau_D}\right)^{8/g}\frac{2L\zeta\left(2-\frac{
16}{g}\right)(2^{2-16/g}-1)}{\pi^2}, & L\ll L_\varphi ,\\
\frac{1}{\sigma R_I^2a^2}\frac{L_\varphi}{\sqrt{2\pi}}\left(\frac{4\tau_{RC}}{\tau_\varphi}\right)^{8/g}\Gamma\left(\frac12-\frac{8}{g}\right), 
& L\gg L_\varphi ,
\end{cases}
\end{equation}
where $\Gamma(x)$ is Euler gamma-function and  $\zeta(x)$ is Riemann zeta-function. The dependence
of $G(0)$ on $L$ at different temperatures is displayed in Fig. \ref{fig3}. At low $T$ it 
shows a pronounced maximum at $L \sim L_\varphi$ which can be conveniently used for experimental analysis
of low temperature dephasing of Cooper pairs in NS systems. 

\begin{figure}[h]
\includegraphics[width=0.99\columnwidth]{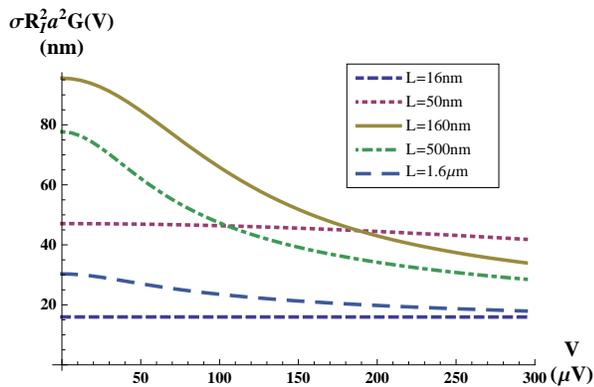}
\caption{(Color online) $G(V)$ at $T=0$ and different values of $L$. The parameters are the same as in Fig. 3.} \label{fig4}
\end{figure}

The same information can also be extracted from the non-linear subgap conductance $G(V)$
which shows the ZBA peak at low voltages \cite{VZK,HN,HN1,Ben,Zai}. At $T \to 0$ and $L\gtrsim L_\varphi$ 
the width of this peak  is roughly determined by $\sim 1/\tau_\varphi$.

In particular, for $L\gg L_\varphi$ and $T=0$ we get
\begin{equation}
G(V)\simeq\frac{1}{\sigma
R_I^2a^2}\frac{L_\varphi}{\sqrt{2\pi}}\left(\frac{4\tau_{RC}}{\tau_\varphi}
\right)^{8/g}{\rm Re}\frac{
\Gamma\left(\frac12-\frac{8}{g}\right)}{ (1+ieV\tau_\varphi)^{1/2-8/g} }
\end{equation}
The non-linear subgap conductance $G(V)$ is depicted in Fig. \ref{fig4} at different values of $L$.

Finally we note that our analysis also allows to determine the subgap conductance for other geometries.
E.g., in 3d case the decoherence effect from spatially uniform fluctuations is negligible \cite{ZGalakt} and
at $T \ll 1/\tau_\varphi$ the dephasing of Cooper pairs in the N-metal is controlled by the second term
in Eq. (\ref{flim}) with $\tau_\varphi \sim \sigma D^{1/2}\tau_c^{3/2} \propto D^3$.

In conclusion, we have demonstrated that electron-electron interactions yield
dephasing of Cooper pairs penetrating from a superconductor into
a diffusive normal metal. At low $T$ this phenomenon imposes fundamental
limitations on the proximity effect in NS hybrids restricting the penetration length
of superconducting correlations into the N-metal to a temperature independent value $L_\varphi$.
This new length scale can be probed by measuring the subgap conductance
in NS systems.

We finally note that our results are qualitatively consistent with experimental observations \cite{Chand}
demonstrating that the low temperature magnetoconductance of NS structures is determined
by phase coherent electron paths with a typical size restricted by the temperature independent 
dephasing length $L_\varphi$  rather than by the thermal length $L_T$ diverging in the low temperature limit. 

This work was supported by the Act 220 of the Russian Government (project 
25) and by RFBR grant 12-02-00520-a.


\begin{thebibliography}{}
\bibitem{FN} This approximation is equivalent to dropping Coulomb blockade effects which are negligible 
in the limit of large $g$ except at exponentially low voltages and temperatures. 
For instance, one can demonstrate \cite{ZGalakt} that for $g \gg 1$ the terms  $\langle \mathcal K^+\mathcal K^-\rangle$
give negligible contribution to Andreev conductance as compared to that of the terms $\langle \mathcal K^+\mathcal K^+\rangle$.
\bibitem{FN0} We do not present a detailed discussion of the Cooperon decay at non-zero $T$ here, since in this regime 
$L_T$ becomes shorter that $L_\varphi$ and, hence, the electron dephasing length becomes irrelevant for the problem under consideration.
\end{thebibliography}
\end{document}